\documentclass[graybox]{svmult}

\usepackage{mathptmx}       % selects Times Roman as basic font
\usepackage{helvet}         % selects Helvetica as sans-serif font
\usepackage{courier}        % selects Courier as typewriter font
\usepackage{type1cm}        % activate if the above 3 fonts are
\usepackage{makeidx}         % allows index generation
\usepackage{graphicx}        % standard LaTeX graphics tool
\usepackage{multicol}        % used for the two-column index
\usepackage[bottom]{footmisc}% places footnotes at page bottom

\makeindex             % used for the subject index
\newcommand{\tar}{\tilde \alpha_r}
\newcommand{\be}{\begin{equation}}
\newcommand{\ee}{\end{equation}}
\newcommand{\ba}{\begin{eqnarray}}
\newcommand{\ea}{\end{eqnarray}}
\newcommand{\ban}{\begin{eqnarray*}}
\newcommand{\ean}{\end{eqnarray*}}
\newcommand{\bml}{\begin{mathletters}}
\newcommand{\eml}{\end{mathletters}}

\def\half{\textstyle{1\over2}}

\def\T0{{\bar{ T_0}}}

\newcommand{\nn}{\nonumber}

\begin{document}

\title*{From Lovelock to Horndeski's generalized scalar tensor theory}
\author{Christos Charmousis}
\institute{Christos Charmousis \at Laboratoire de Physique Th\'eorique (LPT), Univ. Paris-Sud, CNRS UMR 8627, F-91405 Orsay, France \email{christos.charmousis@th.u-psud.fr}}
%\and Name of Second Author \at Name, Address of Institute \email{name@email.address}}
\maketitle

\abstract{We review and discuss some recent progress in Lovelock and Horndeski theories modifying Einstein's General Relativity. Using as our guide the uniqueness properties of these modified gravity theories we then discuss how Kaluza-Klein reduction of Lovelock theory can lead to effective scalar-tensor actions including several important terms of Horndeski theory. We show how this can be put to practical use by mapping analytic black hole solutions of one theory to the other. We then elaborate on the subset of Horndeski theory that has self-tuning properties and review a generic method giving scalar-tensor black hole solutions.}

\section{Introduction}
\label{chasec:1}

General Relativity (GR) is a classical or effective theory of gravity which is based on very solid mathematical and physical foundations. It agrees with overwhelming accuracy local{\footnote{Local distance scales range up to 30 odd astronomical units, size of the solar system, but also size of typical binary pulsar systems. The astronomical unit is a rough earth to sun distance}} observational tests both for weak and strong gravity \cite{chawill} including laboratory tests of Newton's force law. GR, is not only a very successful physical theory. It is theoretically very robust and and as it turns out mathematically a unique metric theory. Indeed if one considers a theory depending on a massless metric and up to its second derivatives endowed with a Levi-Civita connection then, 
\be
\label{chaEH}
S^{\left( 4 \right)}  = \frac{{{m_{Pl}}^4 }}{2}\int {d^4 x\sqrt { - g^{\left( 4 \right)} } \left[ {R - 2\Lambda} \right]} \,,
\ee
is the unique action giving equations of motion of second order in the metric field variable. This theorem, as we will see, is a consequence of Lovelock's theorem \cite{chalovelock} (see also \cite{chalanczos}, \cite{chazumino}). In other words, GR plus a cosmological constant is the unique gravity theory constructed out of a single massless metric  with a Levi-Civita connexion (which is also defined uniquely). This means that any other curvature scalar would necessarily yield either trivial or higher order than second derivatives in the field equations. Higher order than second, $\nabla^2$-derivatives, lead directly to a theory with ghost vacua, clearly an important setback for any classical physical theory  \cite{chaostro}. The only other term evading this problem is the one associated to the cosmological constant. We will encounter consequences of this very shortly.

As we emphasized GR is an {\it effective} metric field theory. As such we expect Einstein's theory to break down at very high energies (strong curvatures) close to the Planck scale,
 $m_{Pl}^2=\frac{1}{16\pi G}$,  where higher order curvature terms can no longer be neglected and are even dominant compared to the leading Einstein-Hilbert term. What is maybe more surprising is that recent cosmological observations, point towards the tantalizing possibility that GR may also be modified at very low energy scales deep in the infra-red \cite{chasuper}. A tiny positive cosmological constant generates an inversely proportional enormous cosmological horizon and can account very simply for such a dark energy component. After all, as we saw in (\ref{chaEH}), it is a mathematically  allowed term in the metric action. However, the difference in between cosmological and local scales corresponds to an enormous number, of magnitude of $10^{15}$, in other words we are very deep in the infra red and physics may well differ from scales where we control gravity observationally. Furthermore, although a cosmological constant provides a phenomenologically correct and economic way to put away the dark energy problem it suffers from a theoretical short-coming, the cosmological constant problem \cite{chaweinberg}. Indeed, from very simple field theory considerations, GR, from its founding strong equivalence principle perceives all forms of matter in time and space including vacuum energy. Vaccum energy gravitates just as does radiation or matter. The cosmological constant, for example, receives zero point energy contributions from each particle species up to the UV cut-off of the relevant QFT. These contribute to the total value of the 
cosmological constant which has to be fine-tuned to almost zero by the arbitrary bare contribution we saw above (\ref{chaEH}).  This fact only gets worse once we realize that phase transitions in the early universe will actually shift this value around, and again each time some miraculous fine-tuning will be required to tune the overall cosmological constant to its tiny but non-zero value we observe today. The "big" cosmological constant problem is precisely how
all these vacuum energies associated to the GUT, SUSY, the standard model etc are fined-tuned each time to zero by an exactly 
opposite in value bare cosmological
 constant $\Lambda_{bare}$ appearing in (\ref{chaEH}) and being the net result of the universe acceleration today.
The unexplained small value of the cosmological constant $\Lambda_{now}$ 
is then an additional two problems to add to the usual "big" cosmological constant problem \cite{chaweinberg}, 
namely, why the cosmological constant is not cancelled exactly to zero and why do we observe it now. In a later section we will see of such an attempt to classically{\footnote{For a interesting proposal tackling the cosmological constant problem including the crucial radiative corrections see, \cite{chakaloper}} evade this problem \cite{chafab4}.

Although this is not really a scientific argument, one can make reference to a historically parallel situation. At the advent of General Relativity, observational evidence pointed towards shortcomings of Newton's gravitational theory in the Òstrong gravityÓ regime. Amongst these was the advance of the perihelion of Mercury, which deviated from Kepler's laws describing planetary motion. As such, the existence of a small planet in an even closer orbit to the sun, Vulcan, was hypothesized. Alternatively, the presence of an unknown substance, aether, was put forward, mediating and slightly modifying the prediction of Kepler's laws to account for observational data. Indeed a simple and slight correction to the established laws of the time could account correctly for the advance of the perihelion. The solution to the puzzle was, however, not as simple or economic as initially considered. In fact, it was only after the theory of GR was put forward that this slight difference was accounted for as, rather, a fundamental modification of gravity theory. As often in physics, a modified physical theory is attained upon reaching a critical energy scale; here, the critical scale in question is the strong gravity field of the sun applied to its closest planet. There is in recent times an observational parallel to the above, in the context of type Ia supernovae explosions, pointing towards an accelerating universe \cite{chasuper}. Friedmann's laws, in order to remain valid, require the addition of an as yet unknown dark energy component, which is the dominant component in the Universe. The addition of a small cosmological constant  gives very good agreement with observational data and is the most economic (in terms of additional degrees of freedom) phenomenological explanation of the acceleration phenomenon. Given, however, the above example, it seems to us important to entertain the following question:  could it be that recent observations are pointing towards a fundamental modification of gravity rather than a modification in the unknown matter sector? Are novel observations indications of a new gravity theory beyond GR? This question is even more compelling since we know that dark matter is so far unaccounted for and in the ultraviolet GR needs to be modified anyway. 
A second important point concerns the predictions and motivation of a modified gravity theory.  Indeed, as we argued above, the initial conditions calling for a modified gravity theory are in order to account on the one hand for the late-time acceleration of the universe and to provide on the other hand a well-defined limit at local scales where the theory at hand should be indistinguishable from GR. This is of course an important and difficult initial step that provides a filter for possible theories under consideration, but this is not all. Since observations can be accounted for by a small cosmological constant put in by hand, one needs to go further in order to make new accurate predictions theoretically. These novel predictions are the real motivation in a modified theory of gravity. Indeed, General Relativity's great successes are not the explanation of the advance in the perihelion of Mercury or its classical limit to Newtonian theory, but rather, completely novel ideas and solutions stemming from the theory itself, such as black holes, Big Bang inflationary theory and so forth. 

So how do we go about modifying such a robust theory such as GR (see the review \cite{chapadilla})? Not surprisingly it is extremely hard both observationally but also theoretically, the windows of modification are rather narrow. This is at the same time fortunate because at the end not there are not too many possibilities left over. In rather loose terms following (\ref{chaEH}) and not breaking some fundamental symmetry like Lorentz invariance (see for example \cite{chablas}), there are four at least routes emanating from a Lagrangian formulation. First, suppose we keep the single massless metric character of the theory. Then inevitably we have to consider higher dimensions. We will show that the relevant theory is then Lovelock theory (see for example \cite{chareviews}). Secondly suppose we stick to 4 space-time dimensions. Then inevitably we consider the existence of additional fields, in other words we add novel gravitational degrees of freedom in 4 dimensional space-time. Here the prototype is scalar-tensor theory and we know its most general form, Horndeski theory \cite{chahorndeski}. We will study basics of this theory here. All the terms present in Horndeski theory have been shown to be originating from Galileons i.e. scalar tensor terms having Gallilean symmetry in flat space-time \cite{chanicolis0} and the latter equivalent theory to Horndeski has been elegantly given for curved space-time in \cite{chadeffayet}. Thirdly we can consider that the elementary particle mediating spin 2 gravity, the graviton, has a finite range of application. In other words it is not a massless field but has some (small) mass. This is the theme of massive gravity \cite{chamassive} which will also be covered in later lectures. Lastly we can consider the possibility of allowing for other geometric constructions such as a differing connexion than that of Levi-Civita. This allows for torsion i.e. non zero parallel transport of scalars (rather than vectors) or first order formalism, Palatini formalism (see for example \cite{chaolmo}). These four directions are not independent of each other in fact often they are related and it is useful to use information from one to the other. We will give such relations during these lectures. We will discuss in fact Lovelock and Horndeski theory and relate the two via the Kaluza-Klein formalism.

Using as our guide uniqueness theorems we will discuss certain elements of Lovelock and Horndeski theory. We will see in what sense these theories are unique. We will focus throughout on recent elements of Lovelock theory that we will be using in relation to Horndeski theory. We will omit some basic properties diverting  the interested reader to \cite{chareviews}. We will then go on to discuss Horndeski theory which is the most general scalar-tensor theory in four dimensions. We will then move on to review some black hole solutions of Lovelock theory and see how, by toroidal Kaluza-Klein reduction we can construct 4 dimensional scalar-tensor black holes. In this way we will establish a clear and practical connection in between Lovelock and Horndeski theory. In the fifth section we will discuss the cosmological constant problem and define a theory which is a subset of Horndeski theory and has interesting self-tuning properties. This theory dubbed fab four \cite{chafab4}, will at least classically provide a partial solution to the big cosmological constant problem. We will then sketch a recent and relatively simple way to obtain black hole solutions in such scalar-tensor theories \cite{chaeugeny}. 

\section{The Lovelock and Horndeski uniqueness theorems}

\subsection{Lovelock theory}

Our purpose in this lecture is to present Lovelock theory in relation to Horndeski theory. To this end it is mostly sufficient to truncate Lovelock theory to what is usually called Einstein-Gauss-Bonnet (EGB) theory in the literature. Unlike the name suggests, this is the 5 or 6 dimensional version of Lovelock theory originally discussed by Lanczos \cite{chalanczos}. 
Let us start with the uniqueness theorem defining Lovelock theory and stick to 6 dimensions in order to fix notation. The 5 dimensional theory is identical.
Consider $\mathcal{L}=\mathcal{L}(\mathcal{M},g,\nabla, \nabla^2)$ where $(\mathcal{M},g)$ is a 6 dimensional locally differentiable Lorentzian manifold without boundary {\footnote{the result depicted here is easily extended to manifolds with boundaries \cite{chareviews}}} and  $\nabla$ is the Levi-Civita metric connexion over $\mathcal{M}$.
The field equations obtained upon metric variation of the action,
 \be
S^{\left(6 \right)}  = \frac{{{M_{(6)}}^4 }}{2}\int {\sqrt { - g^{\left( 6 \right)} } \left[ {R - 2\Lambda  + \alpha \hat G} \right]} \,,
\ee
are unique and admit the following properties:
\begin{itemize}
			\item they depend on a symmetric two-tensor ${\cal E}_{AB}$
			\item the equations of motion are 2$^{\textrm{nd}}$-order PDE' s with respect to the metric field variables
			\item satisfying Bianchi identities.
		\end{itemize}
Here, $M_{(6)}$ is the fundamental mass scale in six-dimensional spacetime,  $\hat G$ is the Gauss-Bonnet density reading,
\be
\label{chagbterm}
\hat G = R_{ABCD} R^{ADCB}  - 4R_{AB} R^{AB}  + R^2 \, ,
\ee
and $\Lambda$ is the cosmological constant. The field equations in vacuum are
\be
{\cal E}_{AB}  = G_{AB}  + \Lambda g_{AB}  + \alpha H_{AB}  = 0\,,
\label{chagbeq}
\ee
where $G_{AB}$ stands for the standard Einstein tensor. Uppercase Latin indices will refer to six-dimensional coordinates whereas greek indices will always refer to 4 dimensional space-time. We have also introduced the Lanczos or second order Lovelock tensor,
\be
\label{chalanczos}
H_{AB}  = \frac{{g_{AB} }}{2}\hat G - 2RR_{AB}  + 4R_{AC} R^C _{\;\;B}  + 4R_{CD} R^{C\;\;D} _{\;\;A\;\;B}  - 2R_{ACDE} R_B ^{\;\;CDE} \,.
\ee
Naturally, the Lanczos tensor is also divergence free, $\nabla^A H_{AB}=0$.
It is important to note that, just like GR in 4 dimensions, i.e. under the same set of hypotheses, EGB theory is the unique and most general metric theory with second order PDE's in 5 or 6 space-time dimensions. This is a non-trivial statement since the terms appearing in the action already contain second order derivatives. In a moment we will see that in higher than 6 dimensions this property is generalized by adding the relevant higher order Lovelock terms. Furthermore in 4 dimensions the tensor (\ref{chalanczos}) is identically zero. Therefore we can note as a prelude that Lovelock theory is the unique massless metric theory in arbitrary dimensions identical to GR with a cosmological constant in 4 dimensional spacetime. 

Before moving on it is useful to discuss some tensorial properties. The Lanczos tensor (\ref{chalanczos}) can be elegantly written (in arbitrary dimension) using the following rank four tensor that will be useful to us later on,
\be \label{chap}
P_{ABCD} = R_{ABCD}+R_{BC}\;g_{AD}-R_{BD}\;g_{AC}-R_{AC}\;g_{BD}+R_{AD}\;g_{BC}+\frac{1}{2} R \;g_{AC} \;g_{BD}-\frac{1}{2} R \;g_{BC} \;g_{AD} ,\ee
as
\be
\label{chaddual}
H_{AB}=-2 P_{ACDE}R_B{ }^{CDE}+\frac{{g_{AB} }}{2}\hat G \, .
\ee
The 4 index tensor $P_{ABCD}$ has several interesting tensorial properties. For a start it is divergence free (in all indices) since Bianchi identities of the curvature tensor are simply written as $\nabla^D P_{ABCD}=0$. It has the same index symmetries as the Riemann curvature tensor. Its bi-tensor obtained by tracing two of its non-consecutive indices yields 
\be
P^{B}{ }_{ACB}=(D-3) G_{AC},
\ee
the Einstein tensor. In fact divergence freedom of the Einstein tensor can be seen to originate from this relation. In a nutshell, one can say that $P_{ABCD}$ is the curvature tensor whose bi-tensor is the Einstein tensor, just as the Ricci tensor is the bi-tensor of the Riemann tensor. A last interesting property is that a metric is an Einstein space, $R_{AB}=\frac{g_{AB}}{D}R$, if and only if $P_{ABCD} = R_{ABCD}$. 

In four dimensions, the $P_{\mu\nu\rho\sigma}$ tensor is even more very special. Indeed it has all the above properties but, additionally it can be pictured in a very similar way to the Faraday tensor in electromagnetism,
\be
\star F_{\mu\nu}=\half \epsilon_{\mu\nu\kappa\rho} \; F^{\kappa\rho}
\ee
In analogy here, $P_{\mu\nu\rho\sigma}$ is a 4 tensor, and coincides with the double dual (i.e. for each pair of indices) of the Riemann tensor defined as, 
\be
P^{\mu\nu}{}_{\rho\sigma} =(^\star R^\star)^{\mu\nu}{}_{\rho\sigma} \doteq -\frac{1}{2} \epsilon^{\rho\sigma\lambda\kappa}\,R_{\lambda \kappa}{ }^{\xi \tau} \, \frac{1}{2} \epsilon_{\xi \tau\mu\nu},
\label{chaddual2}
\ee
where $\epsilon_{\mu\nu\rho\sigma}$ is the rank 4 Levi-Civita tensor. Finally since in 4 dimensions we have that $H_{\mu\nu}=0$ we obtain the Lovelock identity,  
\be
P_{\alpha\nu\rho\sigma}R_\beta{ }^{\nu\rho\sigma}=\frac{{g_{\alpha\beta} }}{4}\hat G
\ee
which will be useful to us later on (see \cite{chaEdgar:2001vv} for extensions).

In order to define the generic Lovelock densities one can use the elegant language of differential forms \cite{chareviews}. Alternatively we take the route taken by Lovelock using the generalized Kronecker delta symbols; the same route taken later on by Horndeski,
\ba
\label{chadelta}
\delta^{A_1...A_h}_{B_1...B_h}&=&\left|
\begin{array}{ccc}
\delta^{A_1}_{B_1} &...& \delta^{A_1}_{B_h} \\
\vdots             &        & \vdots             \\
\delta^{A_h}_{B_1} &...& \delta^{A_h}_{B_h}
\end{array}\right|\\
  &=&h!\delta^{A_1}_{[B_1}...\delta^{A_h}_{B_h]}
\ea
which is antisymmetric in any pair of upper or lower indices. In fact we have $\delta^{A_1...A_h}_{B_1...B_h}=\epsilon_{B_1...B_h} \epsilon^{A_1...A_h}$ with respect to the Levi-Civita symbols. Once this has been digested the Lovelock densities are the complete contraction of the above with the Riemann curvature tensor,
\be
\label{chalovdensities}
L_{(h/2)}=\frac{1}{2^h}\delta^{A_1 A_2...A_h}_{B_1 B_2...B_h}\; R_{A_1 A_2}^{\quad \;B_1 B_2} \;...\; R_{A_{(h-1)} A_h}^{  \qquad \quad B_{(h-1)} B_h}
\ee
As such we can check that $L_{(1)}$ is the Einstein-Hilbert term whereas $L_{(2)}$ is the Gauss-Bonnet combination.
This immediately means that for $h>D$ all Lovelock densities vanish. Therefore the Lovelock Langrangian is given by,
\be
L=\sum_{h=0}^{k} c_h L_h
\ee 
where $k=[(D-1)/2]$. The case $h=D$ is quite special because then the Lovelock density is a topological one.
Indeed we can query what is special about Lovelock densities. The answer lies in differential geometry (see for example \cite{chaspivak}). One can trace the origin of such terms in the early works of Gauss who measuring geodesic distances noted that scalar (Gauss) curvature of two dimensional surfaces depended only on the first fundamental form, in other words the intrinsic metric of the surface and its derivatives. This was the basis of what he called the Egregium theorem; scalar curvature (unlike other extrinsic curvature components) does not depend on the variation of the normal vector field on the surface i.e. on how the surface is embedded in 3 dimensional space. Then later on Euler in his work on surface triangulations noted that 2 dimensional surfaces can be topologically classified by their "Euler" number, $\chi$: $\chi[\mathcal{M}]=2-2h$ where $h$ is the number of topological handles. So one can take an arbitrary surface with no boundary and continuously deform it to a sphere, a torus a double torus and so on{\footnote{When a surface has a boundary an analogous result holds}}. This completely classifies topologically 2 dimensional surfaces. In other words all topological properties of 2 dimensional surfaces can be understood or characterized by their Euler number. Gauss and Bonnet essentially related this topological number to a differentiable geometric quantity, the scalar curvature, resulting in the celebrated relation,
\be
\chi[\mathcal{M}_2]=\frac1{4\pi}\int_{\mathcal{M}} R.
\ee
The Gauss-Bonnet theorem on surfaces has nothing to do with the Gauss-Bonnet term given above (\ref{chagbterm}). For our purposes the above Gauss-Bonnet relation means that the Einstein-Hilbert term is in 2 dimensions is a topological invariant i.e. the Einstein tensor in 2 dimensional space-time is identically zero. This analogy goes through for all Lovelock terms as a corollary to the works of Chern \cite{chachern} who generalized the theorem of Gauss and Bonnet to higher dimensions finding the relevant higher order curvature scalars. For example we have, 
\be
\chi[\mathcal{M}_4]=\frac1{32\pi^2}\int_{\mathcal{M}} \hat G 
\ee
 and thus the Lanczos or Gauss-Bonnet density is a topological invariant in 4 dimensions whose integral is the generalised Euler or Chern topological number. Beware this does not mean that the Gauss-Bonnet scalar is zero or constant in 4 dimensions. It means that the Lanczos density is identically zero $H_{\mu\nu}=0$ as we admitted earlier.  

Dimensionally extending the Chern scalar densities we obtain the Lovelock densities (\ref{chalovdensities}) i.e. just those densities whose variation leads to second order field equations. Any higher order derivatives present in the variation of Lovelock densities conveniently end up as total divergent terms and thus do not contribute to the field equations. In a similar way for example,  in $7$ or $8$ dimensions, the 6-dimensional Euler density will be promoted to a Lovelock density of third power in the curvature tensor and so forth. This explains the nice and unique properties of the Lovelock densities and Lovelock theory in general. For more details the reader can consult \cite{chareviews}.

\subsection{Horndeski theory}

So much for the moment concerning higher dimensional metric theories. In 4 space-time dimensions we know that the unique classical metric theory is GR with a cosmological constant. Hence any 4 dimensional modification of gravity will have to involve some other non-trivial field. The simplest of cases is when this extra field is a scalar. The prototype of  scalar tensor gravity is Brans-Dicke theory \cite{chaBD} which has been studied extensively throughout the years (see \cite{chafarese} and references within). We should note that in the class of scalar-tensor theories fall also other modified gravity theories like $f(R)$ or $f(\hat G)$ which \cite{chasotiriou} are just particular scalar-tensor theories in disguise. Furthermore other interesting GR modifications such as bigravity or massive gravity theories \cite{chamassive} admit scalar tensor theories as particular limits, for example the decoupling limit for massive gravity \cite{chaarkani}. Hence scalar tensor theories are a consistent prototype of GR modification and their important properties are expected in some form,  in other consistent gravity theories. Hence the particular recent interest in scalar-tensor theories concerning modification of gravity. So in this section we reiterate the question: what is the most general scalar tensor theory in 4 dimensional space-time yielding second order field equations? The answer has been given by Horndeski a long-time ago \cite{chahorndeski} but has remained unnoticed  since only recently \cite{chafab4}, and states a similar theorem to that of Lovelock for 4 dimensional scalar-tensor theories. Consider a single  scalar field $\phi$ and a metric $g_{\mu\nu}$ as the gravitational degrees of freedom of some Lorentzian manifold endowed with a Levi-Civita connection. Consider a theory that depends on these degrees of freedom and an arbitrary number of their derivatives,
\be
\label{chahorni}
{\cal L}=
{\cal L}(g_{\mu\nu}, g_{\mu\nu,i_1},...,g_{\mu\nu,i_1...i_p},\phi,\phi_{,i_1},...,\phi_{,i_1...i_q})
\ee
 with $p,q\geq 2$. The finite number of derivatives signifies that we have again an effective theory since we have a finite number of degrees of freedom.
 Here just like in usual Brans Dicke theory we consider that matter couples only to the metric and not to the scalar field thus fixing the metric and the frame as the physical one. In this frame the metric will continue to verify the weak equivalence principle. In a nutshell the metric in question can always be put locally in a normal frame where by definition the Christophel symbols are identically zero. This frame is locally equivalent to an inertial frame. 
The Hornedski action can be written in such a way to involve only second derivatives and reads,
 \ba
\label{chahorndeski}
{\cal L}_H&=& \kappa_1(\phi ,\rho)\delta^{\alpha \beta \gamma} _{\mu\nu\sigma}\nabla^\mu\nabla_\alpha \phi  R_{\beta \gamma} ^{\;\;\;\;\nu\sigma}
           -\frac{4}{3}\kappa_{1,\rho}(\phi ,\rho)\delta^{\alpha \beta \gamma} _{\mu\nu\sigma}\nabla^\mu\nabla_\alpha \phi\nabla^\nu\nabla_\beta \phi\nabla^\sigma\nabla_\gamma \phi \\\nonumber
        &~&+\kappa_3(\phi ,\rho)\delta^{\alpha \beta \gamma} _{\mu\nu\sigma}\nabla_\alpha \phi \nabla^\mu\phi  R_{\beta \gamma} ^{\;\;\;\;\nu\sigma}
           -4\kappa_{3,\rho}(\phi ,\rho)\delta^{\alpha \beta \gamma} _{\mu\nu\sigma}\nabla_\alpha \phi \nabla^\mu\phi \nabla^\nu\nabla_\beta \phi \nabla^\sigma\nabla_\gamma \phi \\\nonumber
        &~&+[F(\phi ,\rho)+2W(\phi )]\delta_{\mu\nu}^{\alpha \beta }R_{\alpha \beta }^{\;\;\;\;\mu\nu}
           -4F(\phi,\rho)_{,\rho}\delta_{\mu\nu}^{\alpha \beta }\nabla_\alpha \phi\nabla^\mu\phi \nabla^\nu\nabla_\beta \phi \\\nonumber
        &~&-3[2F(\phi ,\rho)_{,\phi }+4W(\phi )_{,\phi }+\rho\kappa_8(\phi,\rho)]\nabla_\mu\nabla^\mu\phi 
           +2\kappa_8\delta_{\mu\nu}^{\alpha \beta }\nabla_\alpha \phi \nabla^\mu\phi \nabla^\nu\nabla_\beta \phi \\\nonumber
        &~&+\kappa_9(\phi ,\rho),\\\nonumber
\rho&=&\nabla_\mu\phi \nabla^\mu\phi ,
\ea
%This action represents the most general scale tensor theory that upon variation with respect to the metric and scalar field gives second order field equations satisfying Bianchi identities. 
The action (\ref{chahorndeski}) is rather general and depends on four arbitrary functions  $\kappa_i(\phi,\rho)$, $i=1,3,8,9$  of the scalar field $\phi$ and its kinetic term denoted as $\rho$. Furthermore,
\ba
\label{chaFdef}
F_{,\rho}&=&\kappa_{1,\phi}-\kappa_3-2\rho\kappa_{3,\rho}
\ea
with $W(\phi)$ an arbitrary function of $\phi$, which means we can set it to zero without loss of generality by absorbing it  into a redefinition of $F(\phi, \rho)$. According to Horndeski's theorem, \cite{chahorndeski}, the action  (\ref{chahorndeski}) is the unique{\footnote{In the action one can always add terms that can be written as a total divergence. Therefore the term "unique action" refers to the unique class of equivalence which is in turn defined modulo total divergence terms. In other words two actions are equal if and only if they are in the same class of equivalence or they differ only by a totally divergent term.}} action whose variation with respect to the scalar and metric yields second order field equations and Bianchi identities.
In his original work, Horndeski makes just like Lovelock, systematic use of the anti-symmetric Kronecker deltas (\ref{chadelta}).
The equations of motion are obtained by variation of the metric and scalar field, are parametrized by the arbitrary functions $\kappa_i(\phi,\rho)$ and read respectively, 
\be
{\cal E}^{\mu\nu}=\half T^{\mu\nu}, ~ {\cal E}_\phi=0
\ee
where  $T^{\mu\nu}=\frac{2}{\sqrt{-g}}\frac{\delta S_m}{\delta g_{\mu\nu}}$ is the matter energy-momentum tensor. The tensor ${\cal E}^{\mu\nu}$ is divergent free. A rather more intuitive and economic way of obtaining the Horndeski action is given in terms of the general Galileon covariant action \cite{chadeffayet} and reads, 
\begin{eqnarray}
{\cal L}_{DGSZ}=K(\phi, \rho)-G_3(\phi, \rho) \nabla^2 \phi+G_4 (\phi, \rho)R+G_{4, \rho} \left[ (\nabla^2 \phi)^2-(\nabla_\mu \nabla_\nu \phi)^2 \right]
\nonumber\\
+G_5(\phi, \rho) G_{\mu\nu}\nabla^\mu \nabla^\nu \phi-\frac{G_{5, \rho}}{6} \left[(\nabla^2 \phi)^3-3\nabla^2 \phi (\nabla_\mu \nabla_\nu \phi)^2 +2(\nabla_\mu \nabla_\nu \phi)^3 \right]
 \end{eqnarray}
 In this version it is far easier to recognize subsets of this theory, GR, Brans-Dicke, K-essence etc and to figure out the most common Galileon terms. 
 Again the theory depends on 4 free potentials. 
It was shown in \cite{chaKobayashi:2011nu} that in four dimensions Horndeski's theory is equivalent to the generalised galileon theory with the potentials given by, 
\ba
K&=& \kappa_9+\rho \int^\rho d\rho' \left(\kappa_{8, \phi}-2\kappa_{3, \phi\phi}\right)\\
G_3 &=& 6(F+2W)_{, \phi}+\rho \kappa_8+4\rho  \kappa_{3, \phi}-\int^\rho d\rho' \left(\kappa_{8}-2\kappa_{3, \phi}\right) \\
G_4 &=& 2(F+2W)+2\rho \kappa_3 \\
G_5 &=& -4\kappa_1
\ea
The uniqueness proof by Horndeski is quite technical and can be found in his original paper. Here we simply sketch its important steps. The proof is based on the property relating the metric and scalar field equaltions, 
\be
\label{chaplessis}
\nabla^\mu {\cal E}_{\mu \nu}=\half {\cal E}_\phi \nabla_\nu \phi 
\ee 
This is of course an identity and shows explicitly that the scalar field equation results from the metric equations of motion as a Bianchi identity. Now starting from (\ref{chahorni}) and requiring that ${\cal E}_{\mu \nu}$ and ${\cal E}_\phi $ have second at most derivatives automatically means that this will also have to hold for $\nabla^\mu {\cal E}_{\mu \nu}$. In general if ${\cal E}_{\mu \nu}$ is of second order this is not true for $\nabla^\mu {\cal E}_{\mu \nu}$ but here it is required from (\ref{chaplessis}). So Horndeski starts by finding the most general symmetric, second order 2-tensor $A_{\mu \nu}$ whose divergence $\nabla^\mu A_{\mu \nu}$ is also of second order. This places constraints on the form of $A_{\mu \nu}$ leaving a solution parametrized with 10 free functions. These tensors include of course ${\cal E}_{\mu \nu}$ but not all of them verify (\ref{chaplessis}). Finally then  Horndeski imposes (\ref{chaplessis}) on the former family. This leaves him with 4 free functions at the end giving his final result (\ref{chahorndeski}). 

Now we have at hand the general scalar-tensor and higher dimensional metric gravity framework we will move on to see some of their solutions and how the theories are in fact related in practical terms. We will in particular use known solutions from Lovelock theory in order to construct Horndeski solutions.

\section{Seeking exact solutions in Lovelock theory}
\label{chasec:2}

One of the nice characteristics of Lovelock theory is that despite its additional technical difficulties related to the higher order nature of the theory, certain uniqueness black hole theorems of GR remain valid; at least under some weaker hypotheses. In particular, a generalization of Birkhoff's theorem remains true apart from a case of fine tuning of coupling parameters{\footnote{The special relation between the coupling parameters corresponds to the strong coupling limit of EGB-literally the case where the Gauss-Bonnet term is of maximal relative strength to the Einstein-Hilbert term and gives a very special theory with enhanced symmetries, usually referred to as Chern-Simons theory (see the nice review \cite{chazaneli}). }} \cite{chazegers}. Let us review the higher dimensional version of this result,  for this will lead us to  some relatively simple yet interesting solutions where Lovelock theory even circumvents problems of higher dimensional general relativity. The solutions we will consider will also have a nice application to Galileon/Horndeski theories leading us to black hole solutions for 4 dimensional scalar tensor theories. 

The Birkhoff theorem states that, in four dimensions, any spherically symmetric solution to Einstein's equations in the vacuum is necessarily locally static. In other words there exists a local time like Killing vector. This leads to the celebrated Scharszchild metric as the unique GR solution of spherical symmetry in vacuum. The theorem is not modified when one includes  a negative or positive cosmological constant but the solution itself is slightly more general. Indeed  a negative cosmological constant allows also for exotic horizon topologies of flat or hyperbolic geometry. The general solution of the Einstein field equations with a cosmological constant in $D=4$ dimensions assuming a constant curvature 2-space (rather than a 2-sphere) reads,
\be
\label{chabirk1}
ds^2=-V(r)dt^2+\frac{dr^2}{V(r)}+r^2 \left(\frac{ d\chi^2}{1-\kappa\chi^2}+\chi^2 d\phi^2 \right)
\ee
where the constant $(t,r)$ sections are $2$-dimensional constant curvature spaces parametrized by normalized curvature $\kappa=0,\pm 1$. For linguistic simplicity we will call the surfaces of constant $(t,r)$, horizon sections, preluding the presence of a black hole. The lapse function in (\ref{chabirk1}) reads, $V(r)=\kappa-\frac{\Lambda}{3}r^2-\frac{2M}{r}$. Note then that since the metric is static, zeros of $V=V(r)$ correspond to Killing horizons and exist for $\Lambda<0$ even when $\kappa=0$ or $\kappa=-1$. This can be explicitly checked by going to an Eddington-Finkelstein chart,
\be
\label{chaef}
v=t \pm \int \frac{dr}{V(r)}
\ee
 These black holes are often called topological due to the fact that special identifications have to be made in order for the horizon to be compact \cite{chatopological}. For $\Lambda\geq 0$ only the spherical topologies give regular solutions with the presence of an extra cosmological horizon. So much for 4 dimensional GR with a cosmological constant.

In higher, $D$ dimensional, GR Birkhoff's theorem remains valid not only for constant curvature sections, but also for horizon sections which are Einstein spaces \cite{chagibbons}. Substituting the constant curvature surface of the horizon sections with a $(D-2)$-dimensional Einstein manifold will not alter locally the black hole lapse function and the general solution is static. The structure of space-time locally{\footnote{We will see that when the horizon sections carry non-zero curvature there is a  global change in the topology of the solution related to the presence of a solid angle deficit. This will end up having important consequences that we will discuss in detail later with the solution at hand.}} transverse to the horizon sections is in this way not affected by the details of the internal geometry, as long as the latter continues to be an Einstein space. In particular the horizon structure is the same. To picture this let us take a particular example: consider the, for example, 6 dimensional solution,
\be
\label{chaeuclidean}
ds^2=-V(r)dt^2+\frac{dr^2}{V(r)}+r^2 \left(f(\rho)d\tau^2+\frac{d\rho^2}{f(\rho)}+\rho^2 d\Omega^2_{II} \right)\nn
\ee
where $f(\rho)=1-\frac{\mu}{\rho}$.  Hence the horizon sections in 4 dimensions are given by a Euclidean Schwarzschild black hole obtained by Wick rotating the time coordinate of the original 4 dimensional black hole. The metric (\ref{chaeuclidean}) is a valid 6 dimensional solution since the horizon sections are Ricci flat. On the hand we can consider a second solution of the form,
\be
ds^2=-V(r)dt^2+\frac{dr^2}{V(r)}+r^2 dT^2_{IV}\nn
\ee
with now toroidal horizon sections, i.e. a locally flat 4 dimensional metric. In both cases we have the same lapse function $V(r)=k^2 r^2-\frac{m}{r^3}$ independently whether our horizon is of flat or Euclidean Schwarzschild geometry (which is of course Ricci flat but has non zero Weyl curvature)-Ricci flatness of both the horizon sections means that $\kappa=0$ for the lapse function $V(r)$.
The former exotic black holes often have classical instabilities  \cite{chaGibbons:2002pq} in a similar fashion to those of the black string \cite{chaGregory:1993vy}. In fact black string metrics can be Wick rotated to a subclass of metrics with exotic horizons. The exotic horizon section in this case is nothing but the Euclidean version of 4 dimensional Schwarzschild (as in the example above). We see therefore that in higher dimensional GR a certain kind of degeneracy appears in the possible solutions which are not completely fixed by the symmetries and the field equations. Therefore one could entertain the possibility that the additional unphysical exotic black holes are just an artifact of not considering the full classical gravity theory in higher dimensions. Indeed we will provide clear indications that this is the case at least for 6 dimensional Lovelock theory in the sense that the possible horizon geometries will be seen to be far more constrained \cite{chagleiser} and asymptotically non-trivial.

So how are these results translated in Lovelock theory? In order to answer this question, \cite{chabogdanos}, we start by considering an appropriate anzatz for the metric and stick to $D=6$ dimensions and EGB theory. We have a transverse 2-space, which carries the timelike coordinate $t$ and the radial coordinate $r$, and an internal 4-space, which is going to represent the horizon sections of the possible six-dimensional black holes. The metric of the internal 4 dimensional space we note $h_{\mu\nu}$ and we take to be an arbitrary metric of the internal coordinates $x^\mu, \mu=0,1,2,3$ only. We furthermore impose that the internal and transverse spaces are orthogonal to each other. This is immediately true for GR as a result of the theorem of Frobenius but not true for Lovelock theory. It is an additional assumption we have make in order to make the problem tractable \cite{chabogdanos}. The quite general metric anzatz for which we want to solve the EGB field equations boils down to,
\be
\label{chametric1}
ds^2  = e^{2\nu \left( {t,z} \right)} B\left( {t,z} \right)^{ - 3/4} \left( { - dt^2  + dz^2 } \right) + B\left( {t,z} \right)^{1/2} h^{\left( 4 \right)} _{\mu \nu } \left( x \right)dx^\mu  dx^\nu  \,.
\ee
Using light-cone coordinates,
\be
u = \frac{{t - z}}{{\sqrt 2 }},\quad v = \frac{{t + z}}{{\sqrt 2 }}.
\ee
the metric reads
\be
\label{chabirk2}
ds^2  =  - 2e^{2\nu \left( {u,v} \right)} B\left( {u,v} \right)^{ - 3/4} dudv + B\left( {u,v} \right)^{1/2} h^{\left( 4 \right)} _{\mu \nu } \left( x \right)dx^\mu  dx^\nu  \,.
\ee
We want to solve Lovelock's equations (\ref{chagbeq}) for metric (\ref{chabirk2}).  The key to doing so boils down to the following two equations: the $(u u)$ and $(\upsilon \upsilon)$ equations that literally play the role of integrability conditions for the full system of equations of motion \cite{chabowcock},
\be
\label{chauu}
{\cal E}_{uu}=\frac{2 \nu_{,u} B_{,u}- B_{,uu}}{B} \left[ 1+\alpha \left( B^{-1/2} R^{(4)}+\frac{3}{2} e^{-2\nu} B^{-5/4} B_{,u} B_{,v}  \right) \right]\,,
\ee
\be
\label{chavv}
{\cal E}_{vv}=\frac{2 \nu_{,v} B_{,v}- B_{,vv}}{B} \left[ 1+\alpha \left( B^{-1/2} R^{(4)}+\frac{3}{2} e^{-2\nu} B^{-5/4} B_{,u} B_{,v}  \right) \right].
\ee
The above permit to classify and eventually completely solve the full system of field equations \cite{chabogdanos}. We have three classes of solutions depending on wether the second, the first factor is zero, or again a third class for constant $B$. Here we concentrate on the class of most interest, class II, corresponding to,
\be
\label{chaclass2}
\frac{2 \nu_{,v} B_{,v}- B_{,vv}}{B}=0, \qquad \frac{2 \nu_{,u} B_{,u}- B_{,uu}}{B} =0
\ee
The other classes are degenerate and occur for special relations of couplings only. Most importantly class II solutions are directly connected to GR since (\ref{chaclass2}) is independent of the coupling constant $\alpha$. Solving (\ref{chaclass2}) immediately shows that we have a locally static space-time \cite{chadufaux} and thus a somehow weaker version of Birkhoff's theorem still holds. 

Solving the remaining field equations leads eventually to the metric solution \cite{chagleiser}, \cite{chabogdanos},
\be
\label{chaClass2a}
ds^2  =  - V\left( r \right)dt^2  + \frac{{dr^2 }}{{V\left( r \right)}} + r^2 h^{\left( 4 \right)} _{\mu \nu } \left( x \right)dx^\mu  dx^\nu \,,
\ee
with lapse function,
\be
V(r) = \frac{{R^{\left( 4 \right)} }}{{12}} + \frac{{r^2 }}{{12\alpha }}\left[ {1 \pm \sqrt {1 +\frac{12\alpha \Lambda}{5}  +\frac{{\alpha ^2 \left({R^{(4)}}^2-6\hat G^{(4)} \right)}}{{r^4 }} + 24\frac{{\alpha M}}{{r^5 }}} } \right]\,,
	\label{chapotentialclassII}
\ee
Note first that there are two branches of solutions. This is true generically in EGB theory and results from the higher order nature of the theory \cite{chareviews}. In EGB there are generically two vacua for a given theory{\footnote{In higher order Lovelock theory there are more according to the order of the highest order Lovelock term \cite{chareviews}}}. The upper $'+'$ branch does not have a well-defined GR limit ($\alpha \rightarrow 0$) and turns out to be unstable (for a full recent discussion on stability of EGB vacua see \cite{chapad}). The lower branch is ghost free \cite{chazwie} and is the branch that we will consider from now on omitting the $'+'$ branch. 
The horizon sections, parametrized by the 4 dimensional metric $h^{\left( 4 \right)} _{\mu \nu }$ are constrained by the field equations to be Einstein spaces, 
\be
\label{chaeinstein}
R_{\mu\nu}^{(4)}=\frac{R^{\left( 4 \right)}}{D-2}h^{\left( 4 \right)}_{\mu\nu}
\ee
 But as we can see from the lapse function which has to be a function of the radial variable $r$, a new geometric condition now appears whereupon the horizon quantity ${R^{(4)}}^2-6\hat G^{(4)}$ has to be constant. Combining the two conditions gives,
\be
\label{chagleiserdotti}
C^{\alpha \beta \gamma \mu} C_{\alpha \beta \gamma \nu}=\Theta \delta^\mu_\nu
\ee
where $\Theta$ is a given constant and $C_{\alpha \beta \gamma \nu}$ is the 4 dimensional Weyl tensor associated to $h^{\left( 4 \right)} _{\mu \nu }$.
This is a supplementary condition for EGB theories, the Dotti-Gleiser condition, (\ref{chagleiserdotti}) imposed in addition to the usual Einstein space condition (\ref{chaeinstein}) for higher dimensional general relativity. Clearly then, for EGB theory, the lapse function for the black hole carries a supplementary information particular to the type of horizon section for the black hole solution. For example, the Euclidean Schwarszchild metric is not a legitimate internal metric anymore since it does not verify (\ref{chagleiserdotti}). Both of the conditions (\ref{chaeinstein}) and (\ref{chagleiserdotti}) present a geometric similarity in that we ask for (part of) {\it the curvature tensor to be analogous to the spacetime metric}. The main difference being that the curvature tensor in  (\ref{chagleiserdotti}) is the Weyl tensor and, given its symmetries, it is actually its square which has to be analogous to the spacetime metric. Clearly horizons with $\Theta\neq 0$ will not be homogeneous spaces and not even asymptotically so. Another interesting point is that the Gauss-Bonnet scalar, whose spacetime integral is the Euler characteristic of the horizon, has to be constant for these solutions to be valid.
The Gauss-Bonnet scalar of the internal space then reads $\hat G^{(4)}=4 \Theta+24 \kappa^2$ and the potential \cite{chagleiser}, \cite{chabogdanos},
\be
V(r)=\kappa+\frac{r^2}{12 \alpha} \left(1- \sqrt{1+\frac{12}5\alpha\Lambda - 24 \frac{\alpha^2 \Theta}{r^4} + 24 \frac{\alpha M}{r^5}} \right)\, .
	\label{chapotentialclassIIEinstein}
\ee
since $R^{\left( 4 \right)}=12 \kappa$. For $\Theta=0$, we obtain the black holes first discussed by Boulware and Deser (see \cite{chaBoulware}). In this case since the Weyl curvature is zero the horizon sections are geometries of constant curvature. Taking $\Lambda=0$ we note that these black holes are asymptotically flat and are an extension of the higher dimensional version of the Schwarzschild solution. In fact taking the limit of $\alpha$ small and large $r$ one obtains precisely the latter GR solution. Recently these black holes have been reported to have a spin 2 instability for small enough mass parameter (\cite{chagleiserdotti0}). This result has been extended to Lovelock black holes (\cite{chasoda0}). It is not yet understood what is the physical nature of this "short distance scale instability" and if it is somehow related to thermodynamic instability and quantum Hawking radiation. 

Putting this aside we now want to examine cases of Einstein metrics whose squared Weyl curvature is not zero but constant. This is a special case of Einstein metric and the Dotti-Gleiser condition is much like a supplementary requirement. What is already clear is that any such solution will not be asymptotically "usual" as the fall off the relevant term is $5$ rather than $6$ dimensional. Indeed notice that the $\Theta$-term in (\ref{chapotentialclassIIEinstein}) has a fall off rate of a 5-dimensional Boulware-Deser black hole \cite{chaBoulware} and is therefore dominant over the "usual" mass term contribution, \cite{chamaeda}.  We now investigate a simple example which will have interesting 4 dimensional consequences.

Consider a four-dimensional space which is a product of two 2-spheres, 
\be
ds^2  = \rho _1 ^2 \left( {d\theta _1 ^2  + \sin ^2 \theta _1 d\phi _1 ^2 } \right) + \rho _2 ^2 \left( {d\theta _2 ^2  + \sin ^2 \theta _2 d\phi _2 ^2 } \right)\,,
\ee
where the (dimensionless) radii $\rho_{1}$ and $\rho_{2}$ of the spheres are constant. The entire six-dimensional metric reads,
\be
ds_{(4)}^2  =  - V\left( r \right)dt^2  + \frac{{dr^2 }}{{V\left( r \right)}} + r^2 \rho _1 ^2 \left( {d\theta _1 ^2  + \sin ^2 \theta _1 d\phi _1 ^2 } \right) + r^2 \rho _2 ^2 \left( {d\theta _2 ^2  + \sin ^2 \theta _2 d\phi _2 ^2 } \right)\,,
\label{cha2sphere}
\ee
with lapse function
\be
\label{cha2sphere2}
V\left( r \right) = \frac{{R^{\left( 4 \right)} }}{{12}} + \frac{{r^2 }}{{12\alpha}}\left( {1 - \sqrt {1 - 24k^2 \alpha - 24\Theta \frac{{\alpha ^2 }}{{r^4 }} + 24 \alpha \frac{M}{{r^5 }}} } \right)
\,.
\ee
In order for (\ref{cha2sphere}) to be a solution to the equations of motion the spheres have to be of equal radius, $\rho_{1}=\rho_{2}$. This ensures that (\ref{cha2sphere}) is an Einstein space. The second condition is then immediately verified for a product of 2-spheres. We have $\kappa=\frac1{3\rho_1^2}>0$ and $\Theta=\frac4{3\rho_1^4}$. Note that even when the 2-sphere curvature is normalized to $\rho_1=1$ then $\kappa\neq 1$. A linear redefinition of the $r$ coordinate then shows that the area of the 4 dimensional space is reduced compared to the homogeneous 4-sphere. In other words space-time is asymptotically altered by an overall solid angle deficit. This results in a genuine curvature singularity at $r=0$. Of course when we have $M\neq 0$ there is central curvature singularity at $r=0$ anyway. But, for (\ref{cha2sphere}) the $r=0$ singularity is present {\it even for zero mass whenever $\Theta \neq 0$}! This is not an artefact of EGB theory. In fact it is easy to see, taking the combined limit of $\alpha \rightarrow 0$ and large $r$, that the resulting GR black hole with $V(r)=\frac{{R^{\left( 4 \right)} }}{{12}}+r^2 k^2-\frac{M}{r^3}$ has exactly the same problem at the origin independently of the value of $M$. For $M=0$ the GR solution has a naked singularity at the origin. Note again that the lapse function for higher dimensional GR is the same with the higher dimensional Schwarzschild black hole modulo the horizon curvature term. The zero mass solution is singular at the origin wether we are in GR or Lovelock theory. But for Lovelock theory an interesting effect occurs due to the presence of the $\Theta$ term in the lapse function.

To see this consider for the moment $M=0$ in the lapse function (\ref{cha2sphere}). Then the $\Theta$ term in (\ref{cha2sphere}) is identical to the mass term in the Boulware Deser black hole in 5 dimensions. Therefore, as we know from the Boulware-Deser solution \cite{chaBoulware}    this extra $\Theta$ term generically generates an event horizon cloaking the central $r=0$ singularity as long as $\alpha\neq 0$! In fact the length scale of this event horizon is given by the coupling constant $\alpha\sim length^2$ which we know from string theory effective actions \cite{chatseytlin} is related to the fundamental string tension $\alpha'$. One then can interpret this horizon as a higher order 'quantum' cloak of an otherwise naked singularity present in GR. Details for the horizon structure can be found in \cite{chabogdanos}. We will come back to this solution in order to construct a Galileon black hole.

Let us, before moving on, make some final remarks regarding these solutions.  First we should note that most probably these multiple sphere solutions can be unstable to linear perturbations. It has been shown in GR \cite{chaGibbons:2002pq} that there is a "balloon instability" whereupon one of the spheres wants to deflate with respect to the other. This geometric effect may  remain true in the above EGB version \cite{chagleiser} although the perturbation equations for EGB in lesser symmetry change completely compared to GR. It is also probable that this instability may be stabilized by the inclusion of a magnetic field in the relevant solution \cite{chayannis}. Secondly we should note that the above construction involving multiples of equal radius spheres, can be undertaken in arbitrary even dimensional spacetime as long as we truncate Lovelock theory to EGB. If one considers higher order Lovelock terms it is not known under what geometric conditions the horizon sections will be admissible. One may expect a higher order condition of the type (\ref{chagleiserdotti}) in third and higher curvature order...  We expect the horizon sections to be more and more constrained as higher order Lovelock terms come into play. At the same time since horizon sections will be of higher dimension this will allow for a richer geometry. This is also an open question. Finally, putting it all together we have arrived to the following result concerning EGB theory: {\it given the anzatz (\ref{chametric1}), the only asymptotically flat solution of $6$ dimensional EGB theory with zero cosmological constant, is the Boulware Deser solution \cite{chaBoulware}}. This is because whenever $\Theta\neq 0$ the solution is not asymptotically flat for 6 dimensional space-time. Therefore we can deduce that EGB theory is very similar in this aspect to 4 dimensional GR lifting the degeneracy present in higher dimensional GR due to the additional elegant geometric condition (\ref{chagleiserdotti}).

\section{From Lovelock to Horndeski theory: Kaluza-Klein reduction}
\label{chasec:3}

In order to apply higher dimensional Lovelock theory to cosmology or gravity in 4 dimensional space-time one needs some means of approach to 4 dimensional gravity. There are at least two routes, braneworlds and Kaluza Klein reduction. In the recent past Lovelock theory had an important implication in the braneworld paradigm \cite{charandall}. Braneworlds consist of higher dimensional spacetimes endowed with a distributional brane where standard matter is localized. The idea "inspired" in rather loose terms from string theory, is that gravity perceives all the space-time dimensions while matter is localized on a 4 dimensional braneworld. Since the set-up involves junction or matching conditions an essential feature is the number of extra dimensions, namely codimension, yielding for example a wall or string type of defect. There is a long literature of articles on the subject treating codimension one \cite{chareviews} and codimension two braneworlds (see \cite{chacod2} and references within) involving respectively five and six dimensional EGB theory. In particular Lovelock theory permits, due to the generalized junction conditions \cite{chazegers2}, well defined codimension two braneworld cosmology \cite{chakofinas}. This leads to important consequences since in GR one cannot consider distributional sources for cosmological symmetry and codimension 2.  Again, the richer structure of Lovelock theory permits solutions with distributional sources not available in higher dimensional GR. For more details on these aspects see \cite{chareviews} and \cite{chacod2} and references within. Here, we will focus on the more classical Kaluza-Klein compactification since it  will give us a direct connection to higher order scalar tensor terms, found in Galileon/Horndeski theory. It will also provide a way to obtain exact black hole solutions \cite{chablaise2}. 

It has been known since a long time \cite{chahoyssen} that a consistent Kaluza Klein reduction of Lovelock theory gives a scalar-tensor theory with higher order derivatives, but crucially, with second order equations of motion. In this sense many of the Galileon terms discussed later on were known from previous work on Kaluza Klein compactifications and braneworlds \cite{chaKK}. This is the direction we will take here. The most generic of Kaluza-Klein reduction to 4 dimensions has recently been given in the nice paper of \cite{chaVanAcoleyen:2011mj}. There it has been shown that only up to the third order Lovelock terms contribute to the Kaluza-Klein compactification in 4 dimensions. Here we will concentrate on EGB theory i.e. up to second order Lovelock theory. We will consider the simplest consistent toroidal compactification giving rise to one extra scalar degree of freedom.  

Start by taking $D$-dimensional Einstein Gauss-Bonnet theory which is the 5 or 6-dimensional Lovelock theory truncated to arbitrary dimension. The arbitrary dimension $D$ will be important when we end up promoting dimension from a positive integer to a real parameter once we have undertaken a consistent Kaluza-Klein reduction. We have the EGB action with a cosmological constant,
\be
	S = \frac1{16\pi G_N}\int d^{D}x\,\sqrt{-g}\left[-2\Lambda +R+\hat G \right]
\label{chaGBAction2}
\ee
Consider now the simplest but consistent diagonal reduction along some arbitrary $n$-dimensional internal curved space $\tilde{\mathbf  K}$. We aim to reduce this theory down to $4$ space-time dimensions with $D=4+n$:
\be
	d s^2_{(4+n)}=d \bar s^2_{(4)} + e^{\phi} d \tilde K^2_{(n)}
	\label{chaKKGalileon}\,.
\ee
This particular frame is chosen in such a way as so there is no conformal factor of $\phi$ in front of the $4$-dimensional metric. As such the asymptotic character (i.e. radial fall off) of a Lovelock $D$ dimensional solution will be similar to the 4 dimensional one. All terms with a tilde refer to the curved $n$-dimensional internal space, while terms with a bar refer to the $(4)$-dimensional space-time.  
One can show for the given metric Anzatz that the KK reduction for $n$ arbitrary is consistent, \emph{i.e.} that the reduced equations of motion are derived from the reduced action \cite{chablaise}.
This reduction is therefore generalised  in the manner defined in \cite{chaKanitscheider:2009as,chaGouteraux:2011qh}. The important result of this is that the integer $n$ corresponding to the compact Kaluza-Klein space can be analytically continued to a real parameter of the reduced action. Naturally $n$ corresponds to a dimension only for $n$ integer. The solutions from the four dimensional point of view are still solutions of the resulting effective action for arbitrary $n$. The 4 dimensional effective action reads after integrating out the internal space,
\ba\label{chaCurvedGBGalileonAction}
%\begin{split}
	\bar S_{(4)}&=& \int d^{4}x\,\sqrt{-\bar g}\,e^{\frac n2\phi}\left\{\bar R -2\Lambda +\alpha \bar G+\frac n4(n-1)\partial\phi^2 -\alpha n(n-1)\bar G^{\mu\nu}\partial_\mu\phi\partial_\nu\phi \right.\nonumber \\
			&-&\frac\alpha4n(n-1)(n-2)\partial\phi^2\nabla^2\phi+\frac\alpha{16}n(n-1)^2(n-2)\left(\partial\phi^2\right)^2 \nonumber\\
			&+&\left.e^{-\phi}\tilde R\left[1+\alpha \bar R+\alpha 4(n-2)(n-3)\partial\phi^2\right]+\alpha\tilde G e^{-2\phi}\right\}\,,
%\end{split}
\ea
For $\alpha=0$ this effective action is just the usual toroidal KK effective action. The higher order Gauss-Bonnet term gives rise to several higher order scalar-tensor Galileon (or equivalently Horndeski) terms, \cite{chanicolis0,chaDeffayet:2009wt,chadeffayet,chaDeffayet:2011gz} with very particular potentials. 
The Galileon field $\phi$ can then simply be understood to be the scalar field parametrising the volume of the internal space. 

Indeed, apart from the usual lower order terms appearing in standard Kaluza Klein compactification of Einstein dilaton theories, we see the emergence of several higher order terms. For a start we have the 4 dimensional Gauss Bonnet term $\bar G$ which will contribute to the scalar field variation although it is a topological term for 4 dimensional GR.  Secondly we have $\bar G^{\mu\nu}\partial_\mu\phi\partial_\nu\phi$ involving the coupling of the Einstein tensor with the kinetic term. Here, rather than metric-scalar interaction, as for the standard kinetic term of $\phi$ we have a curvature-scalar interaction which we will see has very interesting consequences in the forthcoming section. This term has equations of motion of second order essentially due to the divergence free property of the Einstein tensor $G_{\mu \nu}$. For example if one considers $\bar R^{\mu\nu}\partial_\mu\phi\partial_\nu\phi$ this is not true. It  has also shift symmetry in the scalar field typical of certain Galileon terms. Furthermore we have, what is often called  the DGP term,  $\partial\phi^2\nabla^2 \phi$ appearing in the decoupling limit of the DGP model \cite{chaarkani} and then the standard Galileon term $\left(\partial\phi^2\right)^2$ which are also shift symmetric in $\phi$. The last line in the effective action takes part only for a curved internal space in the face of Ricci and Gauss Bonnet curvature. 
Reducing from the EGB action yields terms up to quartic order in derivatives (either of the metric or the scalar, or a mixed combination of the two). Reducing higher order Lovelock densities yields terms with a higher number of derivatives. A typical example is the higher order permissible curvature-scalar interaction,
\be
\label{chapaul}
P^{\mu\nu\alpha \beta} \nabla_\mu \phi \nabla_\alpha \phi \nabla_\nu \nabla_\beta \phi 
\ee
 which involves six derivatives and one can show  \cite{chafab4} originates from the Kaluza-Klein reduction of the third order Lovelock density \cite{chaVanAcoleyen:2011mj}. Again the reader will recollect the divergence freedom of the double dual tensor giving second order field equations. Taking for example $R^{\mu\nu\alpha \beta} \nabla_\mu \phi \nabla_\alpha \phi \nabla_\nu \nabla_\beta \phi $ would fail this Galileon property.

Although this effective action is very complex, and its field equations even more so,
it is "simple" to generate solutions for the above (\ref{chaCurvedGBGalileonAction}) in 4 dimensions \cite{chablaise}. One starts from a convenient Lovelock solution in $D$ dimensions. Since we want the 4 dimensional solution to have, at least locally, spherical horizon sections we have to consider a solution where the $(D-2)$ dimensional horizon sections are $(D-2)/2$-products of two spheres. This is precisely the extension of the 6 dimensional solution we discussed in the previous section (\ref{cha2sphere}) generalized to arbitrary dimensions \cite{chablaise}. The solution reads,
\ba
	d \bar s^2_{(4)} &=& -V(R) d t^2 + \frac{d R^2}{V(R)}+ \frac{R^2}{n+1} d \bar K^2_{(2)}\,, \label{chaGalileonMetric}\\
	V(R)&=& \kappa+\frac{R^2}{\tar}\left[1\mp\sqrt{1-\frac{2\tar}{l^2}-\frac{2\tar^2 \kappa^2}{(n-1)R^4}+\frac{4 \tar m}{R^{3+n}}}\right], \label{chaGalBHPot}\\
	\tar &=&  2\alpha n (n+1), \qquad \frac{1}{\ell^2}=\frac{-2\Lambda}{(n+2)(n+3)} \qquad\\
	e^\phi&=&\frac{R^2}{n+1}\,, \label{chaGalPhi}
\ea
Here, $n$ is the dimension of the internal space minus one 2-sphere in other words, $n=D-4$. This is the higher dimensional interpretation of $n$ but once the solution is written out we simply take $n$ an arbitrary real number and (\ref{chaGalileonMetric}) is still an exact solution and $n$ parametrizes the theory. In our notation here $\kappa=0,1,-1$ is the normalised horizon curvature and we have redefined for this section the constants $\tar$ and $\ell$. Taking carefully the $\tar\rightarrow 0$ limit, gives a standard Einstein dilaton solution with a Liouville potential \cite{chablaise2}. Set $\Lambda=0$, $\kappa=1$ and let us start by making some qualititative remarks describing properties of the solution without entering into technical details. Note that, taking carefully the $n=0$ limit switches off the scalar field and the higher-derivative corrections, and we obtain pure GR in (\ref{chaCurvedGBGalileonAction}) and a Schwarzschild black hole (\ref{chaGalileonMetric}). This is particularly interesting since the scalar-tensor solution given above for arbitrary $n$ is a continuous deformation of the Schwarzschild solution. When $n$ is in the neighborhood of zero we are closest to the GR black hole. As we hinted in the previous section the topology of the solution is not that of GR. Indeed the warp factor of the 2-sphere, in (\ref{chaGalileonMetric}), is recovered only at $n=0$, i.e. the GR limit. Otherwise the area of the reduced spherical horizon is given by $\frac{4\pi R^2}{n+1}$ rather than the 2-sphere area, $4\pi R^2$.  This is again a solid deficit angle (and not a conical deficit angle) the same one we encountered for the Lovelock solution in the previous section. As stressed in the previous sections this will give, at $R=0$, a true curvature singularity even if $m=0$. For large $R$, we have a spacetime metric very similar to that of a gravitational monopole, \cite{chaBarriola:1989hx}.  Expanding \ref{chaGalBHPot} for small $\tar$ and large $R$ gives,
\be
V(R)=1+\frac{\tar }{(n-1)R^2}-\frac{2m}{R^{n+1}}+...
\ee
This solution is reminiscent of a RN black hole solution where the role of the electric charge is undertaken by the leading horizon curvature correction in $\tar$. This is the particular $\Theta$ term we discussed in the previous section. This term dominates the mass term close to the horizon and for $n<1$. Note that it can be of negative sign depending on the value of $n$ and $\tar$. The further we are from $n=0$, the GR limit, the further we deviate from a standard four-dimensional radial fall-off. 
The first important question we want to deal with is the central curvature singularity at $R=0$, which is due to the solid deficit angle and is present even if $m=0$. Also note that whenever the square root in the lapse function (\ref{chaGalBHPot}) is zero we also have a branch singularity which is also a dangerous curvature singularity. Setting $m=0$, we find that  for $-1<n<1$ and $\tar>0$ the singularity at $R=0$ is covered by an event horizon created by the higher-order curvature correction. In its absence ($\tar=0$), this solution would have been singular... The UV (small $R$) behaviour of the solution is therefore regularised by the presence of the higher-order terms. If $n>1$ or $n<-1$, then $\tar<0$ is needed in order to preserve the event horizon. The remaining cases are singular.

Now let us switch on the mass, $m \neq 0$. Whenever $\tar>0$, we have a single event horizon. When $-1<n<1$, there is no branch singularity however small $m$ is. On the contrary, when $n>1$, the mass is bounded from below in order to avoid a branch singularity:
\be
m> \left(\frac{2}{n+3} \right)^{\frac{n+3}{4}} \frac{\tar^{\frac{n+1}{2}}}{n-1}\,.
\ee
When $n<-1$, the solution is also a black hole but the mass term is not falling off at infinity. The region of most immediate interest is whenever $n$ is small but not zero.

The black hole properties are rather different for $\tar<0$. When $-1<n<1$, there is an inner and an outer event horizon as long as the following condition is fulfilled:
\be
\left(\frac{1}{2} \right)^{\frac{n+1}{2}}<\frac{m(1-n)}{|\tar|^{\frac{n+1}{2}}}<\left(\frac{2}{n+3} \right)^{\frac{n+3}{4}}.
\ee
When $n>1$, a single event horizon exists, covering a single branch singularity with $R_s<R_h$.

Overall we can say that the KK solution given here has an interesting horizon structure and presents again a quantum cloaking of an otherwise Einstein-Dilaton singular solution. It is however not of ordinary asymptotics bifurcating in this way the no hair paradigm for Galileons \cite{chanicolis}. In the last section we will see a way to construct asymptotically ordinary solutions with fake black hole hair.

\section{Self-tuning and the fab 4}
\label{chasec:5}

As we saw in the previous sections Horndeski or Galileon theory encompasses all the possible (single) scalar tensor terms one can consider in order for the equations of motion to be of second order. This is an essential requirement for a well-defined classical modification of gravity \cite{chaostro}. In this section we will question which of the terms in scalar-tensor theory have "self-tuning" properties. Self-tuning is a rather simple and quite old idea with application to the cosmological constant problem. The basic principle consists of finding solutions for flat (or possibly  maximally symmetric vacua) of some gravity action endowed with a bulk cosmological constant, independently of the value of the cosmological constant in the action. In order for self-tuning (and not fine tuning) to be effective the cosmological constant should not be fixed with respect to any of the coupling constants in the gravitational action. The idea then is that the cosmological constant is absorbed by a dynamical solution involving the non-trivial scalar field without affecting the gravitational background. This can only be a "partial" solution to the cosmological constant problem since radiative corrections will destabilize this vacuum solution beyond a certain energy scale, the cutoff of the effective gravity theory. It is however an interesting first step especially since no theories were known, before \cite{chafab4}, to have such a property without some hidden effective fine tuning of the action coupling constants as for example in codimension one braneworld models (see for example \cite{chakaloper3}). We should note that recently there has been considerable progress on protecting the cosmological constant from standard model radiative corrections \cite{chakaloper2} and we refer the interested reader to this article for the model in question which interestingly is a rather minimal extension of GR. Rather, for our purposes, having at hand the general scalar tensor theory we will formulate the following question: is there a subset of Horndeski theory with self-tuning properties? The answer is affirmative as shown in \cite{chafab4}, yielding a rather simple and neat geometrical theory which was dubbed by the authors as Fab 4 theory. We start by presenting the theory and then give a specific self tuning solution which elegantly and non-technically gives the general idea. We close the section by showing a simple method to obtain regular black hole solutions in fab 4 and Horndeski theory, independently of self-tuning. 

The Fab 4 potentials make up the most general scalar-tensor theory capable of self-tuning. They are given by the following geometric terms,
  \begin{eqnarray}
\label{chaeq:john}
{\cal L}_{john} &=& \sqrt{-g} V_{john}(\phi)G^{\mu\nu} \nabla_\mu\phi \nabla_\nu \phi \\
\label{eq:paul}
{\cal L}_{paul} &=&\sqrt{-g}V_{paul}(\phi)   P^{\mu\nu\alpha \beta} \nabla_\mu \phi \nabla_\alpha \phi \nabla_\nu \nabla_\beta \phi \\
\label{eq:george}
{\cal L}_{george} &=&\sqrt{-g}V_{george}(\phi) R \\
\label{eq:ringo}
{\cal L}_{ringo} &=& \sqrt{-g}V_{ringo}(\phi) \hat G
\end{eqnarray}
where $R$ is the Ricci scalar, $G_{\mu\nu}$ is the Einstein tensor, $P_{\mu\nu\alpha \beta}$ is the double dual of the Riemann tensor (\ref{chaddual}), $\hat G=R^{\mu\nu \alpha \beta} R_{\mu\nu \alpha \beta}-4R^{\mu\nu}R_{\mu\nu}+R^2$ is the Gauss-Bonnet combination. As we saw in the previous section all of these terms with particular potentials appear in Kaluza-Klein reduction of higher order Lovelock terms. Self tuning solutions exist for any of these potentials as long as either $\{V_{john}\}\neq 0$ or , $\{V_{paul}\}\neq 0$ or $\{V_{george}\}$ are not constant. Note that this constraint means that GR in accordance to Weinberg's no-go theorem \cite{chaweinberg} does not have self-tuning solutions. Also $V_{ringo}$ cannot self-tune but does not spoil self-tuning, i.e. it cannot self-tune without (a little) help from his friends-hence the unfortunate name. Also note that taking $\{V_{george}\}= constant$ as for GR with $\{V_{john}\}\neq 0$ suffices for example to have a self-tuning theory. In fact pure GR does not exclude self-tuning of the theory as long as another non-trivial fab 4 term is present. This is also very interesting from a phenomenological point of view. We will see in what follows how all these facts come about. 

The fab 4 terms are related to particular functionals of the Horndeski potentials,
\ba
\kappa_1 &=&2V_{ringo}'(\phi)\left[1+\frac{1}{2}\ln(|\rho|)\right]-\frac{3}{8}V_{paul}(\phi)\rho \label{chak1}\\
\kappa_3 &=& V_{ringo}''(\phi)\ln(|\rho|)-\frac{1}{8}V_{paul}'(\phi)\rho-\frac{1}{4}V_{john}(\phi)\left[1-\ln(|\rho|)\right] \\
\kappa_8 &=&\frac{1}{2}V_{john}'(\phi)\ln(|\rho|),\\
\kappa_9 &=& -\rho_\Lambda^{bare}-3V_{george}''(\phi)\rho
\\
F+2W &=& \frac{1}{2}V_{george}(\phi)-\frac{1}{4} V_{john}(\phi)\rho\ln(|\rho|) \label{chaf2w}
\ea
Notice that the self-tuning constraints fix completely the dependence on the kinetic term $\rho$. Notice also that the Fab 4 terms are scalar interactions with space-time curvature. No pure potential or kinetic terms are allowed for self-tuning. Again, we will see why their form has to be so special. 

Weinberg`s no-go theorem tells us that our vacuum solution must not be Poincar\'e invariant \cite{chaweinberg}.
Hence if we consider cosmological symmetry with a time dependent background, the scalar field has to depend non-trivially on the time coordinate breaking Poincar\'e invariance for flat space-time.
The self-tuning filter defining the self-tuning property and thus the form of Fab 4 terms is  as follows:
\begin{itemize}
\item Fab 4 terms admit locally a Minkowski vacuum for {\it any} value of the net bulk cosmological constant
\item this remains true before and after any phase transition in time where the cosmological constant jumps instantaneously by a finite amount. The scalar field will have to be able to change accordingly in order to accommodate the novel value without affecting the flat space-time background.
\item Fab 4 terms permit non-trivial cosmologies i.e. does not self-tune for any other matter backgrounds other than vacuum energy.
\end{itemize}
The last condition ensures that Minkowski space is not the only cosmological solution available, something that is certainly required by observation. The idea is that the cosmological field equations should be dynamical, with the Minkowski solution corresponding to some sort of fixed point. In other words, once we are on a Minkowski solution, we stay there -- otherwise we evolve to it dynamically \cite{chasaffin}. This last statement would indicate that the self-tuning vacuum is an attractive fixed point. Mathematically self-tuning under these conditions, and especially the second, translates to a junction condition problem where the metric is regular and $C^2$ whereas the second derivative of the scalar field contains Dirac distribution terms.
%Last important point is that all fab 4 terms with specific potentials result from higher dimensional Kaluza Klein reduction of Lovelock theory. Indeed $\{{\cal L}_{john}\}$, $\{{\cal L}_{george}\}$ and $\{{\cal L}_{ringo}\}$ originate as we saw from the Kaluza-Klein reduction of the 5 dimensional Gauss-Bonnet term. The only term that does obey this rule is $\{{\cal L}_{paul}\}$ as it involves 6 derivatives and it originates from the third order Lovelock term.

The full equations of motion are given by,
\ba
&& {\cal E}^{\mu \nu}_{john}+{\cal E}^{\mu \nu}_{paul}+{\cal E}^{\mu \nu}_{george}+{\cal E}^{\mu \nu}_{ringo}=\half T^{\mu \nu}  \\
&& {\cal E}^\phi_{john}+{\cal E}^\phi_{paul}+{\cal E}^\phi_{george}+{\cal E}^\phi_{ringo}=0
\ea
We have included the cosmological constant  in the energy momentum tensor $T^{\mu \nu}$.
The contribution of each term from variation of the metric is given by 
\ba
&&{\cal E}^{\eta \epsilon}_{john} = \frac{1}{2}V_{john}(\rho G^{\eta \epsilon}-2 P^{\eta \mu \epsilon \nu}  \nabla_\mu \phi \nabla_\nu \phi  )+\nonumber \\
&&+\half g^{\epsilon \theta}\delta^{\eta \alpha \beta}_{\theta \mu\nu} \nabla ^\mu  (\sqrt{V_{john}} \nabla_\alpha \phi) \nabla^\nu ( \sqrt{V_{john}} \nabla_\beta \phi ) \\
&& {\cal E}^{\eta \epsilon}_{paul} =\frac{3}{2} P^{\eta \mu \epsilon \nu} \rho V_{paul}^{2/3}  \nabla_{\mu}  \left(V_{paul}^{1/3}  \nabla_\nu \phi \right)  \nonumber \\
&&+\half g^{\epsilon \theta} \delta^{\eta \alpha \beta \gamma}_{\theta \mu\nu\sigma}  \nabla^\mu \left(V_{paul}^{1/3} \nabla_\alpha\phi\right) \nabla^\nu  \left(V_{paul}^{1/3} \nabla_\beta\phi\right) \nabla^\sigma \left(V_{paul}^{1/3} \nabla_\gamma\phi\right) \\
&& {\cal E}_{george}^{\eta \epsilon}= V_{george} G^{\eta \epsilon}-(\nabla^\eta \nabla^\epsilon - g^{\eta \epsilon}  \nabla^\rho \nabla_\rho)V_{george} \\
&& {\cal E}^{\eta \epsilon}_{ringo}=-4 P^{\eta \mu \epsilon \nu} \nabla_\mu \nabla_\nu V_{ringo}
\ea
and from variation of the scalar by 
\ba
&&{\cal E}^\phi_{john} = 2 \sqrt{V_{john}} \nabla_\mu (\sqrt{V_{john}} \nabla_\nu \phi) G^{\mu\nu} \\
&&{\cal E}^\phi_{paul} = 3 V_{paul}^{1/3}  \nabla_\mu \left(V_{paul}^{1/3} \nabla_\alpha\phi\right) \nabla_\nu  \left(V_{paul}^{1/3} \nabla_\beta\phi\right)P^{\mu\nu \alpha \beta} -\frac{3}{8} V_{paul} \rho \hat G \\
&&{\cal E}^\phi_{george} = -V_{george}' R \\
&&{\cal E}^\phi_{ringo} = -V_{ringo}' \hat G
\ea
Notice that the scalar equation of motion vanishes identically for flat space-time. This necessary condition can be traced back to the distributional origin of the scalar field and strongly characterizes these terms. Indeed note that a canonical kinetic term for the scalar is disqualified from self-tuning because there is no matter source to account for the distributional part of the scalar field. This is why fab 4 terms represent curvature-scalar interactions: so that their scalar field equations are redundant for the self-tuning background in question.

Instead of going through the detailed derivation of the self-tuning terms in Horndeski theory we will rather look at a simple cosmological example in order to see how self-tuning works in practice. For the details we refer the interested reader to the original papers, \cite{chafab4}.
 
In order to evade Weinberg's no-go argument concerning the cosmological constant we have to break Poincar\'e invariance for the scalar field. As such we consider a time-dependent scalar field and the FRW family of cosmological metrics of the form,
\be
\label{eq:cosmometric}
ds^2  = -dT^2+a^2(t)\gamma_{ij} dx^i dx^j
\ee
where $\gamma_{ij}$  is the metric on the unit plane ($\kappa=0$), sphere ($\kappa=1$) or hyperboloid ($\kappa=-1$).The Friedmann equation reads  ${\cal H}=-\rho_\Lambda^{bare}$ as we are supposing only vacuum energy to be present, 
\be \label{chahamff}
{\cal H}={\cal H}_{john}+{\cal H}_{paul}+{\cal H}_{george}+{\cal H}_{ringo}+\rho_\Lambda^{bare}
\ee
and
\ba
&&{\cal H}_{john}=3V_{john}(\phi)\dot\phi^2\left(3H^2+\frac{\kappa}{a^2}\right) \nonumber\\
&&{\cal H}_{paul}=-3V_{paul}(\phi)\dot\phi^3H\left(5H^2+3\frac{\kappa}{a^2}\right) \nonumber\\
&&{\cal H}_{george}=-6V_{george}(\phi)\left[\left(H^2+\frac{\kappa}{a^2}\right)+H\dot\phi \frac{V'_{george}}{V_{george}}\right]\qquad \nonumber\\
&&{\cal H}_{ringo}=-24V'_{ringo}(\phi)\dot\phi H\left(H^2+\frac{\kappa}{a^2}\right) \nonumber
\ea
Self-tuning requires a flat space-time solution and a time dependent non-trivial scalar field whenever $\rho_m=\rho_{\Lambda}$ and for all $\Lambda$. Flat space in cosmological coordinates is given for a hyperbolic slicing $\kappa=-1$ with $a(T)=T$ and $H=1/T$. This is Milne space-time, the cosmological slicing of flat Minkowski space-time.
Therefore, plugging $H^2=-\kappa/a^2$ into (\ref{chahamff}), we immediately see that 
\be \label{chacond1}
V_{john}(\phi)(\dot\phi H)^2+V_{Paul}(\phi)(\dot\phi H)^3-V'_{george}(\phi)(\dot\phi H)+\rho_\Lambda=0
\ee
Here, we immediately see that {\it ringo} or a constant {\it george} do not spoil self-tuning but require necessarily another non-trivial fab 4 term.
Indeed we see that the scalar field $\phi$ is given locally (in space and time) with respect to the arbitrary bulk value of the cosmological constant. This is again an essential condition. For since the scalar field equation is redundant and the space-time metric given,  the Friedmann equation has to fix the scalar field dynamically i.e. with respect to its derivative. Hence the first condition means that the Friedmann equation is not trivial; it depends on $\dot{\phi}$. 
Furthermore, the scalar equation of motion is actually redundant for flat space-time. This is important for otherwise under an abrupt change of the cosmological constant the scalar derivative could not be discontinuous disallowing self-tuning. This is the implementation of the second condition. Indeed the scalar equation $E_\phi=0$, where 
\be \label{ephiff}
E_\phi={ E}_{john}+{ E}_{paul}+{ E}_{george}+{ E}_{ringo}
\ee
and 
\ba
&&{ E}_{john}= 6{d \over dt}\left[a^3V_{john}(\phi)\dot{\phi}\Delta_2\right]  - 3a^3V_{john}'(\phi)\dot\phi^2\Delta_2
 \nonumber\\
&&{E}_{paul}= -9{d \over dt}\left[a^3V_{paul}(\phi)\dot\phi^2H\Delta_2\right]  +3a^3V_{paul}'(\phi)\dot\phi^3H\Delta_2
 \nonumber\\
&&{ E}_{george}= -6{d \over dt}\left[a^3V_{george}'(\phi)\Delta_1\right]  +6a^3V_{george}''(\phi)\dot\phi \Delta_1 
+6a^3V_{george}'(\phi)\Delta_1^2  \nonumber\\
&&{ E}_{ringo} = -24 V'_{ringo}(\phi) {d \over dt}\left[a^3\left(\frac{\kappa}{a^2}\Delta_1 +\frac{1}{3}  \Delta_3 \right) \right]  \nonumber
\ea
with operator 
\be
\Delta_n=H^n-\left(\frac{\sqrt{-\kappa}}{a}\right)^n
\ee
vanishes on shell  for $n>0$.  However, we should note that the third condition is implemented by the fact that the full scalar equation of motion should {\it not be independent} of $\ddot a$. This ensures that the self-tuning solution can be evolved to dynamically, and allows for a non-trivial cosmology. The second Friedmann equation results from the scalar and 1st Friedmann equation as a Bianchi identity. 

In order to explicitly show a self-tuning solution consider some particularly simple potentials that can be obtained by Taylor expansion on $\phi$.
\ba
&& V_{john} = C_j, \qquad V_{paul}=C_p,  \\
&& V_{george}= C_g + C_g^1\, \phi \,, \qquad V_{ringo}=C_r + C_r^1 \,\phi + C_r^2 \,\phi^2 \,,
\ea
This Taylor expansion corresponds to a slow varying late time scalar filed. Since (\ref{chacond1}) is homogeneous in $\dot{\phi}H$ 
it is quite easy to see that
\be
\label{chaselftune}
\phi = \phi_0 + \phi_1 T^2 \,,
\ee
is a solution where $\phi_0$ and $\phi_1$ are constants, 
with
\be
\label{chacond2}
- C_g^1 \, \phi_1
+ 2\, C_j \phi_1^2
- 4\, C_p \phi_1^3
+ \frac{\rho_\Lambda}{6} =0 \,,
\ee
Therefore for arbitrary $\Lambda$ there exists $\phi_1$ satisfying locally (\ref{chacond2}) without fine tuning of the potentials, here $C_{fab4}$. If $\Lambda$ jumps to a different value then so can do $\phi_1$ and this corresponds to a discontinuous scalar field $\dot{\phi}$. The same mechanism occurs for arbitrary potentials, of course there the solution is more complex.  An interesting question now arises: is it possible that self-tuning solutions exist for other vacuum metrics of the theory. Could we for example have Fab 4 with a cosmological constant and find a self-tuning vacuum black hole, in other words a black hole solution than rather than de-Sitter have flat space-time asymptotics. This is still an open problem for the theory, although a self-tuning solution has been recently found in the literature with a remnant effective cosmological constant \cite{chaeugeny}.

Let us now move on into the direction of exact solutions,  describing a method which will give black hole solutions in this theory \cite{chaeugeny}. Let us for simplicity consider two of the Fab 4 terms namely John and George and let us also consider their potentials to be constants. We have therefore the action,
\begin{equation}\label{chaaction2}
S = \int d^4x \sqrt{-g}\left[\zeta R +\beta G^{\mu\nu}\partial_\mu\phi \partial_\nu\phi  \right],
\end{equation}
and here notice we have not included a cosmological constant.
The relevant coupling constants are now $\zeta$ and $\beta$ and as a result the above action is shift-symmetric for the scalar field $\phi$. According to what we described above, this  theory is a self-tuning theory for flat space-time as long as 
$\beta\neq 0$ had we had a cosmological constant in the action. The metric field equations read,
\begin{eqnarray}\label{chaeomg}
\zeta G_{\mu\nu} +\frac{\beta}2 \left[ (\partial\phi)^2G_{\mu\nu} + 2 P_{\mu\alpha\nu\beta} \nabla^\alpha\phi \nabla^\beta\phi \right. \nonumber \\
	 \left.   +  g_{\mu\alpha}\delta^{\alpha\rho\sigma}_{\nu\gamma\delta}\nabla^\gamma\nabla_\rho\phi \nabla^\delta\nabla_\sigma\phi \right]
	=0,
\end{eqnarray} 
where $P_{\alpha\beta\mu\nu}$ is the double dual of the Riemann tensor (\ref{chaddual}).
The $\phi$ equation of motion can be rewritten in the form of a current conservation, as a consequence of the shift symmetry of 
the action,
\begin{equation}\label{chaeomJ}
 \nabla_\mu J^\mu =0,\;\; J^\mu = \beta G^{\mu\nu}  \partial_\nu\phi.
\end{equation}
Note that (\ref{chaeomJ}) contains a part of the metric field equations, namely that originating from the Einstein-Hilbert term. We now consider a spherically symmetric Anzatz
\begin{equation}\label{chametric3}
	ds^2 = -h(r)dt^2 + \frac{dr^2}{f(r)} + r^2 d\Omega^2.
\end{equation}
where $f(r)$, $h(r)$ are to be determined from the field equations. 

Let us make a slight pause in order to make connection with the flat self-tuning solution we exposed previously. In the (\ref{chametric3}) system of coordinates, Milne space-time is given as,
\be
T=\sqrt{t^2-r^2},\qquad \coth X=\frac{t}{r}
\ee
and thus we note from (\ref{chaselftune}) that the self tuning solution we depicted previously (\ref{chaselftune}) is given by $\phi(t,r)=\phi_0 + \phi_1 (t^2-r^2)$. The scalar field therefore in this coordinate chart is a radial, time dependent function. Therefore any self-tuning black hole solution will have to have a time and radially dependent scalar. 

Although we do not find a self-tuning solution for the forthcoming example (we have taken $\Lambda=0$) we consider the Anzatz,
\begin{equation}\label{chacondition2} 
\beta G^{rr}=0, \qquad  \phi(t,r) = q\, t + \psi(r).
\end{equation}
involving a linear time dependence in the scalar field{\footnote{Clearly, had we been seeking a self-tuning solution in the presence of an arbitrary cosmological constant this linear anzatz would not do. We know rather that there must be at large distance a $t^2$ dependence on the scalar field. This unfortunately renders the field equations $t$-dependent and the system cannot admit a non zero mass solution. In other words a self-tuning black hole would have to be part of a radiating space-time. Again this is an open problem.}}. Notice from the field equations (\ref{chaeomg}) that due to shift symmetry no time derivatives are present, the equations of motion are ODE's.
This condition (\ref{chacondition2}) solves not only the scalar but also the $(tr)$-metric equation which is not trivial due to time dependence of the scalar field $\phi$. Therefore (\ref{chacondition2}) is a valid anzatz rendering the whole system integrable. Indeed the remaining equations are solved for,
with $ f  =h = 1- \frac{\mu}{r},$
whereas the scalar field is not trivial and reads,
\begin{equation}\label{chaphi1}
	\phi_{\pm}=qt\pm q\mu\left[2\sqrt{\frac{r}\mu}+ \log\frac{\sqrt{r}-\sqrt{\mu}}{\sqrt{r}+\sqrt{\mu}}\right]+\phi_0
\end{equation}
The regularity of the metric and the scalar field at the horizon can be
conveniently checked by use of the generalized Eddington-Finkelstein coordinates, with the advanced time coordinate, $v$,
\begin{equation}\label{chav}
	v = t + \int ( fh )^{-1/2} dr.
\end{equation}
One finds from (\ref{chametric3}) and (\ref{chav}),
\begin{equation}\label{metricEF}
	ds^2 = - h dv^2 +2\sqrt{h/f}\, dv dr + r^2 d\Omega^2.
\end{equation}
One can explicitly check that the solution (\ref{chaphi1}) with the plus sign does not diverge on the future horizon (whereas the solution with the minus sign is regular on the past horizon). Indeed the transformation (\ref{chav}) reads,
$
v=t+r+\mu\log(r/\mu-1),
$
and using (\ref{chaphi1}) one finds,
\begin{equation}
	\phi_+ = q\left[v-r + 2\sqrt{\mu r} - 2\mu \log\left(\sqrt{\frac{r}{\mu}}+1\right)\right]+const,
\end{equation}
which is manifestly regular at the horizon, $r=\mu$. This is therefore a regular GR black hole with a non-trivial scalar field which is also regular at the horizon. This method can be applied in differing Gallileon contexts yielding relatively simple and well-defined black hole solutions \cite{chatsoukalas}. It seems that the linear time dependence of the scalar field, its shift symmetry and the presence of higher order terms is capital to the presence of regular black hole solutions. Indeed if there is no linear time-dependence then the scalar field can present singular behavior at the horizon and solutions are not asymptotically flat \cite{chaother}. We can re-iterate the Anzatz (\ref{chacondition2}) roughly as long as the Galileon scalar equation of motion gives the metric field equation of the lower order term. In other words gravitational terms go in pairs, as here in our example, the Einstein-Hilbert and the John term. One can show that a similar property holds for Ringo and Paul terms of the Fab 4. Indeed one can show that the scalar equation associated to Paul, $P^{\mu\nu\alpha \beta} \nabla_\mu \phi \nabla_\alpha \phi \nabla_\nu \nabla_\beta \phi$ with $V_{paul}=constant$ gives the metric field equations of $\phi \hat G$. Note also that the latter is also invariant under shift symmetry. This method bifurcates the no-hair arguments in \cite{chanicolis} (see \cite{chaeugeny} and \cite{chazhou}).

\section{Conclusions} 

In this lecture we have studied certain aspects of Lovelock and Horndeski theory that have been discussed very recently in the literature of modified gravity theories. The former theory, as we saw is the general metric theory of massless gravity in arbitrary dimensions and with a Levi-Civita connexion, whereas the latter is the general scalar-tensor theory in 4 dimensional space-time, again using a Levi-Civita connexion. Lovelock theory, is GR with a cosmological constant in 4 dimensions whereas Horndeski theory is GR once the scalar field is frozen. In this sense and given their unique properties the two theories are essential and very general examples of modified gravity theories. General because, for example,  Horndeski theory includes known and widely studied $F(R)$ or $F(\hat G)$ theories. General also since part of Horndeski theory is a limit of  other fundamental modified gravity theories such as massive gravity \cite{chamassive} in its decoupling limit \cite{chaarkani}. We saw that Lovelock and Horndeski theories are explicitly related via Kaluza-Klein reduction and one can map solutions from one theory to the other. This permitted to find analytic black hole solutions in Horndeski theory for the first time \cite{chablaise}. We then moved on to discuss a subset of Horndeski theory which has self-tuning properties. This particular theory consisting of 4 scalar-curvature interaction terms has been dubbed Fab 4 \cite{chafab4}. Although Fab 4 does not present a full solution of the cosmological problem since it does not account for radiative corrections \cite{chakaloper}, the theory itself has some very interesting integrability properties giving for the first time scalar-tensor black holes with regular scalar field on the black hole horizon. The method described briefly here is quite powerful since it can be applied in differing gravitational theories of the Galileon type or even with bi-scalar tensor theories \cite{chatsoukalas}. We have depicted very recent ongoing research directions in these fields which have numerous open problems. We hope that these notes will help in tackling some of those in the recent future.  

\begin{acknowledgement}
I am very happy to acknowledge numerous interesting discussions with my colleagues during the course of the Aegean summer school. I am also indebted to E Papantonopoulos for the very effective and smooth organisation of the school. I am very happy to thank Eugeny Babichev for numerous comments as well as Stanley Deser for remarks on the first version of this paper. I also thank the CERN theory group for hosting me during the final stages of this work
\end{acknowledgement}
%
%\section*{Appendix}
%\addcontentsline{toc}{section}{Appendix}
%
%

%\input{referenc}

\end{document}